\documentstyle[preprint,aps]{revtex}

\begin{document}

{\tighten
\preprint{\vbox{\hbox{CALT--68--1935}\hbox{FERMILAB--PUB--94--132--T}
\hbox{JHU--TIPAC--940006}\hbox{UCSD/PTH 94--06}}}

\title{Another Source of Baryons in $B$ Meson Decays}

\author{Isard Dunietz and Peter S.~Cooper}
\address{Fermi National Accelerator Laboratory, P.O.~Box 500,
Batavia, Illinois 60510}
\author{Adam F.~Falk\thanks{On leave from The Johns Hopkins
University, Baltimore, Maryland}}
\address{Department of Physics, University of California,~San Diego,
La Jolla, California 92093}
\author{Mark B.~Wise}
\address{California Institute of Technology, Pasadena, California
91125}

\date{May 20, 1994}

\maketitle
\begin{abstract}
It is usually assumed that the production of baryons in $B$ meson decays
is induced primarily by the quark level process $b\to c\bar ud$, where the
charm quark hadronizes into a charmed baryon.  With this assumption, the
$\Lambda_c$ momentum spectrum would indicate that the transition
$B\to\Lambda_c X$ is dominated by multi-body $B$ decays.  However, a
closer examination of the momentum spectrum reveals that the mass
$m_X$ against which the $\Lambda_c$ is recoiling almost always satisfies
$m_X\agt m_{\Xi_c}$.  This fact leads us to examine the hypothesis that
the production of charmed baryons in $B$ decays is in fact dominated by
the underlying transition $b\to c\bar cs$, and is seen primarily in modes
with two charmed baryons in the final state.  We propose a number of
tests of this hypothesis.  If this mechanism is indeed important in
baryon production, then there are interesting consequences and
applications, including potentially important implications for the
``charm deficit'' in $B$ decays.
\end{abstract}

}

The interpretation of data on the production of charmed baryons in the
weak decay of $B$ mesons often involves significant model-dependence.  In
particular, it consistently has been assumed in experimental analyses
that baryon production arises predominantly from the quark-level process
$b\to c\bar ud$, where the charm quark fragments to a $\Lambda_c$ or
$\Sigma_c$, which is in turn observed in the cascade decay to a
$\Lambda$~\cite{argus,cleo,zoeller,BHP}.  In this letter, we will suggest
that this may in fact not be the case, that rather, the dominant
quark-level process for charmed baryon production is $b\to c\bar cs$. 
This process is usually neglected, because of the phase space suppression
arising from the mass of the additional charm quark.  We will present
circumstantial evidence that the $b\to c\bar cs$ process actually
contributes significantly to the production of charmed baryons, and
propose a more stringent test of our hypothesis which makes use of
baryon-lepton sign correlations.  If this indeed turns out to be the case,
there are a number of interesting theoretical and experimental
consequences, which we will discuss.

The only charmed baryons which have so far been reconstructed in $B$
decays are the $\Lambda_c$ and $\Sigma_c$, which is observed in its decay
to $\Lambda_c\pi$.  Since final states are included with their charge
conjugates to improve the statistics~\cite{argus,cleo,zoeller}, it is not
known whether a given $\Lambda_c$ actually comes from the decay of a $B$ or
a $\overline B$.  However, under the usual assumption that the $\Lambda_c$
is produced directly in the decay of a $\overline B$ meson to a single
charmed hadron, the data exhibit a curious feature.  As pointed out in
Refs.~\cite{argus,cleo,zoeller}, there is absolutely no evidence for
two-body decays of the form $\overline B\to\Lambda_c X$.  Such evidence
would come from the momentum spectrum of the $\Lambda_c$.  We display
the most recent CLEO data in Fig.~\ref{cleodata}, which is taken from
Ref.~\cite{zoeller}.  The spectrum is clearly much too soft to be
consistent with two-body decays.  If one fits the spectrum to $\overline
B\to\Lambda_c\overline N(n\pi)$ (where $N$ is a nucleon), then one has to
take $n\ge3$~\cite{cleo,zoeller}.

In fact, the higher-statistics CLEO study~\cite{cleo,zoeller} is
consistent with finding very few $\Lambda_c$'s with momentum
$P_{\Lambda_c}\agt 1.5\,\text{GeV}$.  This is equivalent to a strong
statement about the invariant mass $m_X$ of the hadronic state against
which the $\Lambda_c$ is recoiling, namely $m_X\agt
2.3\,\text{GeV}\approx m_{\Lambda_c}$.  (In fact, the binned data are not
inconsistent with the even stronger condition $m_X\agt m_{\Xi_c}$.) This
is most puzzling, if one believes that the production of $\Lambda_c$'s is
induced by the quark-level transition $b\to c\bar ud$, leading to
$\overline B\to\Lambda_c X$.  One would need to posit a mechanism for
suppressing those final states $X$ with invariant mass $m_p\le
m_X\alt m_{\Lambda_c}$.  

These facts lead us to the hypothesis that the production of
charmed baryons in $B$ meson decays is dominated not by the transition
$b\to c\bar ud$ but by $b\to c\bar cs$.  In
contrast to $b\to c\bar ud$, this process can yield naturally the
$\Lambda_c$ momentum spectrum which is observed.  We illustrate this in
Fig.~\ref{sample}, where we plot the predicted momentum spectrum under the
fairly generic assumption that $\Lambda_c$'s are produced equally in the
two-body modes $\overline\Xi_c\Lambda_c$, $\overline\Xi_c'\Lambda_c$,
$\overline\Xi_c\Sigma_c$ and $\overline\Xi_c'\Sigma_c$.  Here two
charmed baryons are produced per $B$ decay, for example via the quark
diagrams shown in Fig.~\ref{quark}. In Fig.~\ref{sample}, the smearing due
to the small boost of the $B$ meson in the $\Upsilon(4S)$ rest frame has
been included.  The $\Sigma_c$ is seen in its cascade decay to
$\Lambda_c$, while the $\Xi_c$ is too light to decay strongly and hence cannot yield a $\Lambda_c$.  By
the $\Xi_c'$, we mean the spin-${1\over2}$ $SU(3)$ 6 state similar to the
$\Xi_c$, which is a $\overline 3$ under $SU(3)$.  It is the strange
analogue of the $\Sigma_c$, and its mass splitting from the $\Xi_c$ has
been measured to be $95\,\text{MeV}$~\cite{simon}.  We stress that we
present this plot simply to illustrate how naturally the data can be
reproduced by the assumption that $\Lambda_c$'s are produced in $B$ decay
via $b\to c\bar cs$, rather than in $\overline B$ decay via $b\to c\bar
ud$.  This simple model fails to account for the approximately $20\%$ of
$\Lambda_c$'s which have momenta below $0.55\,\text{GeV}$, which must come
from the decays of higher charmed baryon resonances or from many-body
decays.

We note that the $b\to c\bar cs$ transition cannot actually saturate the
production of charmed baryons in $B$ decays, because CLEO has recently
observed the exclusive mode $\overline B\to\Lambda_c\,\overline
p\,\pi^+\pi^-$ at the $0.2\%$ level, while obtaining tight upper limits on
$\overline B\to\Lambda_c\,\overline p\,(n\pi)$, for
$n=1,\ldots,4$~\cite{ross}.  The observed mode constitutes a tiny $4\%$
fraction of the $\Lambda_c$ yield in $B$ decays.  Since nonperturbative QCD
is involved, there is no firm theoretical calculation of the relative
strengths of baryon production via the $b\to c\bar cs$ and $b\to c\bar ud$
transitions, although various model estimates exist~\cite{models}.

Of course, while the evidence in Figs.~\ref{cleodata} and \ref{sample} is
appealing, it is clearly somewhat circumstantial.  A more stringent test
of our hypothesis can be constructed by analyzing correlations between
charmed baryons from one $B$ and the sign of a hard lepton produced by the
weak decay of the other $B$ in the event.  With appropriate cuts, the sign
of the lepton can be used to tag the parent of the charmed baryon as a $B$
or $\overline B$; for example, a hard $\ell^+$ arising from $\overline b$
decay on the other side of the event indicates that the charmed baryon
came from the decay of  a $b$ quark.  Such a study has already been
performed by CLEO for $\Lambda\ell^\pm$ correlations~\cite{cleo}.  One must
be careful to compensate for the effects of $B-\overline B$
mixing.\footnote{This point is discussed in detail in Ref.~\cite{CD},
where it is pointed out that this has not always been done correctly in
the past.}

For example, let us consider $\Lambda_c\ell^\pm$ and $\Xi_c\ell^\pm$ sign
correlations.  If $\Lambda_c$'s are produced only via the transition
$b\to c\bar ud$, then we expect to observe the correlation
$\Lambda_c\ell^+$.  If instead they are produced via $b\to c\bar cs$, then
we expect to find $\Lambda_c\ell^-$.  (This is strictly true only
in the momentum range $P_{\Lambda_c}\ge0.87\,\text{GeV}$.  Below this
momentum, the correlations may partially be spoiled by the presence of a
$\overline\Lambda_c\Lambda_c\overline KX$ final state, where the
$\Lambda_c\overline K$ comes, for example, from the decay of a highly
excited $\Xi_c^{(r)}$ resonance.)  Both the $b\to c\bar cs$ and the $b\to
c\bar ud$ mechanisms predict a $\Xi_c\ell^+$ correlation, while
$\Xi_c\ell^-$ correlations should come only from $b\to c\bar cs$.

It is useful to assemble the information which may be gained from these
correlations into a single unified test of our hypothesis. 
Unfortunately, this cannot be done without introducing a certain amount
of model-dependence, but we will make it as minimal, and as explicit, as
possible.  We consider four mechanisms for the production of charmed
baryons in $\overline B$ decay, corresponding to the quark-level
transitions $b\to c\bar ud$, $b\to c\bar cs$, $b\to c\bar us$ and $b\to
c\bar cd$.  The last two modes are Cabibbo-supressed, but we include them
for completeness.  We might na\"{\i}vely expect them to contribute at
the level of five to ten percent of the Cabibbo-allowed modes.  We neglect
the production of charmed baryons in semileptonic $B$ decays, which is
expected to be small.  Let the notation $B_{\bar ud}$ denote that part of
the branching ratio of $Br(B\to\;\text{baryons})$ which comes from $b\to
c\bar ud$, and define $B_{\bar cs}$, $B_{\bar cd}$ and $B_{\bar us}$
analogously.  We also denote by $R_{H_c\ell^\pm}\equiv
N_{H_c\ell^\pm}/N_{\text{tagged}}$ the yield of charmed hadrons $H_c$
correlated with hard charged leptons $\ell^\pm$, divided by the total
number of lepton-tagged $B\overline B$ events. We assume that $B-\overline
B$ mixing has been corrected for, and, of course, acceptance and detection
efficiencies have been included.  

We need to make some assumption about the relative probability of
producing $s\bar s$ pairs during the fragmentation process, relative to
$u\bar u$ or $d\bar d$ pairs.  Although this could in principle depend on
the particular kinematics of each decay, we will model it by a single
probability $p$, such that for $p=0$ no $s\bar s$ pairs are
produced, and for $p=1$ we have exact $SU(3)$ symmetry in the
fragmentation process.  Unfortunately, we must also make the
dynamical assumption that if a decay is not two-body, then all the quarks
present immediately after the decay of the $b$ materialize in charmed
hadrons, if possible.  For example, we assume that if the underlying
transition is $b\to c\bar ud$, that the charmed baryon is of the
form $cdq$, where $q\bar q$ is produced during fragmentation.  This
assumption is probably not important in the $b\to c\bar cs$ and
$b\to c\bar cd$ channels, where we suspect from the evidence given above
that the decays are primarily two-body, but it is more worrisome for
final states with only one charmed baryon.  Of course, if such states in
fact contribute only minimally to charmed baryon production (as we
suggest), then the assumption is not so dangerous.  Finally, there will
be a small contamination, for example, from the decays of highly excited
charmed baryon resonances, such as $\Xi_c^{(r)}\to\Lambda_c\overline K$,
$\Sigma_c\overline K$, $D\Lambda$, $D\Sigma$, $D^+_s\Xi$, or
$\Lambda_c^{(r)}$,$\Sigma_c^{(r)}\to Dp$, $\Xi_c K$.

We consider five charmed baryon-lepton sign correlations: 
$\Lambda_c\ell^\pm$, $\Xi_c\ell^\pm$, and $\Omega_c\ell^+$.  Assuming
that the fragmentation to baryons in the ground state $SU(3)$ $\overline
3$ and $6$ is preferred, and with $B-\overline B$ mixing removed, we
find
\begin{eqnarray}
   R_{\Omega_c\ell^+} &=& {p\over2+p}\,\bigg(B_{\bar cs}+B_{\bar
     us}\bigg)\,,\cr
   R_{\Xi_c\ell^+} &=& {2\over2+p}\,\bigg(B_{\bar cs}+B_{\bar us}\bigg)
     +{p\over2+p}\bigg(\,B_{\bar ud}+B_{\bar cd}\bigg)\,,\cr
   R_{\Xi_c\ell^-} &=& {p\over2+p}\,\bigg(B_{\bar cs}+B_{\bar
     cd}\bigg)\,,\cr
   R_{\Lambda_c\ell^+} &=& {2\over2+p}\,\bigg(B_{\bar ud}+B_{\bar
     cd}\bigg)\,,\cr
   R_{\Lambda_c\ell^-} &=& {2\over2+p}\,\bigg(B_{\bar cs}+B_{\bar
     cd}\bigg)\,.\nonumber
\end{eqnarray}
Recall that $B_{\bar cd}$ and $B_{\bar us}$ are Cabibbo-suppressed and
expected to be small, so these equations contain more cross-checks
than may appear at first glance.  Our prediction is that the data will
indicate $B_{\bar cs}\gg B_{\bar ud}$.

Another simple test of our hypothesis is to look for
$\Lambda_c\overline\Lambda$ correlations, which will follow from $b\to
c\bar cs$ if the branching ratio for $\Xi_c\to\Lambda X$ is
significant.  By contrast, the $b\to c\bar ud$ process will result in
$\Lambda_c\overline p$ correlations instead.  Of course, the best test
would be to reconstruct fully the exclusive modes
$B\to\Lambda_c\overline\Xi_c$, $B\to\Sigma_c\overline\Xi_c$, and so
forth.  Now that more than a thousand $\Lambda_c$'s have been
reconstructed, it should become feasible to search for such final states. 
Finally, we note that if charm-anticharm two-body decays dominate
inclusive baryon production in $B$ decays, then the decay daughters, such
as $p$, $\Lambda$, $\Xi$ and $\Sigma$, will show a characteristic momentum
dependence different from that predicted by the $b\to c\bar ud$
mechanism.  As the data on momentum spectra improve, it should become
possible to discriminate between the various production mechanisms. 

If our hypothesis holds up under further scrutiny, there are interesting
theoretical and experimental consequences.  First, it would indicate
that the inclusive charm yield from $B$ decays to baryons has been
seriously underestimated.  This would help resolve the ``charm
deficit'', which is the apparent problem that the number $n_c$ of charm
quarks observed per $B$ decay is closer to $1.00\pm0.07$ than to the
expectation based on phase space,
$n_c\approx1.15$~\cite{BHP}.\footnote{The experimental result uses the
branching ratio $Br(B\to D_s^\pm X)\approx8\%$.  A recent measurement of
this quantity is somewhat larger, $Br(B\to D_s^\pm X)=
(12.24\pm0.51\pm0.89)\%$~\cite{browder}.  Including this result would
increase $n_c$ by 0.04.}  In fact, the problem is more serious, because a
theoretical analysis of the semileptonic branching ratio of the $B$ meson
suggests that $n_c$ is {\it larger\/} than na\"{\i}vely expected, closer to
$1.3$~\cite{SVbaff,DFW}.  

The inclusive branching fraction of $B$ mesons to charmed baryons comes
from the measurement of~\cite{argus,cleo,zoeller}
$$
   \bigg[ Br(B\to\Lambda_c X)+Br(\overline B\to\Lambda_c X)\bigg]
   \,Br(\Lambda_c\to pK^-\pi^+)\,.
$$
The most accurate measurement of this quantity to date is from
CLEO~\cite{zoeller}, who report $(0.181\pm0.022\pm0.024)\%$. 
Coincidentally, Refs.~\cite{cleo} and~\cite{CD} both obtain a $\Lambda_c$
yield of $6\%$ per bottom meson, using very different assumptions.  While
Ref.~\cite{cleo} assumes that the $b\to c\bar ud$ mechanism governs
$\Lambda_c$ production, Ref.~\cite{CD} uses current data under the
assumption of $b\to c\bar cs$ dominance.  Those $\Lambda_c$'s which are
produced via $b\to c\bar cs$, rather than via $b\to c\bar ud$, contribute
two charm quarks, rather than one, to the inclusive charm yield.  Hence,
if charmed baryon production is indeed dominated by $b\to c\bar cs$, then
there is a new contribution to $n_c$ of about $0.06$, or maybe more.  From
a theoretical point of view this would be most welcome.

Our hypothesis must also be considered in the light of the
$\Lambda\ell^\pm$ correlations which have already been observed.  If
one follows the usual assumption that the predominant source of
$\Lambda$'s is $\Lambda_c$'s, then the $b\to c\bar ud$ mechanism would
result in a significant $\Lambda\ell^+$ correlation, which already has
been seen by CLEO~\cite{cleo}.  This correlation can be explained in
the $b\to c\bar cs$ mechanism only if it turns out that the branching
ratio $Br(\Xi_c\to\Lambda X)$ is much larger than $Br(\Lambda_c\to\Lambda
X)$.  There also exists a measurement of inclusive $\Xi^-$ production
in $B$ decays, $Br(B\to\Xi^-X)+Br(\overline B\to\Xi_-X)=
0.27\%$~\cite{cleo,argus2}, which can only be consistent with our
hypothesis if $Br(\Xi_c\to\Xi^-X)+Br(\Lambda_c\to\Xi^-X)$ is small.

However, if charmed baryon production is indeed dominated by the $b\to
c\bar cs$ transition, then much of the current ARGUS and CLEO data on
charmed baryons must be reinterpreted.  A thorough analysis, which is
beyond the scope of this letter, will be presented in Ref.~\cite{CD}. 
There it is found that a consistent alternative picture of the
production and decay of charmed baryons emerges, in which all existing
experimental constraints are satisfied.  In this scenario, the dominant
source of the $\Lambda$'s which have been observed in $B$ decays is the
decay of $\Xi_c$ rather than of $\Lambda_c$.

Finally, we point out that our hypothesis would imply that $\Xi_c$
and $\Omega_c$ baryons are being produced at $B$ factories at a rate far
greater than has heretofore been appreciated.  This raises the exciting
possibility that their properties may be studied in great detail.

We are grateful to T.E.~Browder, D.H.~Miller, W.R.~Ross, M.M.~Zoeller and
the CLEO Collaboration for informing us of their latest results, and for
giving us Fig.~\ref{cleodata}.  I.D.~thanks J.D.~Lewis for insightful
comments.  This work was supported by the Department of Energy under
Grants  DOE-FG03-90ER40546, DE-AC03-81ER40050 and DE-AC02-76CHO3000.

\begin{figure}
\caption{The weighted average of the shape of the $\Lambda^+_c$ momentum
spectrum in $B$ decays compared (a) to the same spectrum derived from
CLEO~1.5 data and (b) to shapes derived from Monte Carlo simulation of the
decays $\overline{B} \to \Lambda^+_c\overline{N}(m\pi)$, with $m =
0,\ldots, 4$ and $N$ denoting $p$ or $n$. All simulated curves have been
normalized to data, with the exception of the case $m=0$, where the
normalization is arbitrary.  The figure is taken from Ref.~[3].}
\label{cleodata}
\end{figure}

\begin{figure}
\caption{The momentum spectrum $P_{\Lambda_c}$, under the assumption
that $\Lambda_c$'s are produced from $B$ decays equally in the two-body
modes $\overline\Xi_c\Lambda_c$, $\overline\Xi_c'\Lambda_c$,
$\overline\Xi_c\Sigma_c\to\overline\Xi_c\Lambda_c\pi$ and
$\overline\Xi_c'\Sigma_c\to\overline\Xi_c'\Lambda_c\pi$.  The random
boost of the $B$ relative to the $\Upsilon(4S)$ has been accounted
for.  The data sample consists of 4000 $\Lambda_c$'s.}
\label{sample}
\end{figure}

\begin{figure}
\caption{Quark diagrams for the production of two charmed baryons from
the decay of a bottom meson.}
\label{quark}
\end{figure}


\begin{references}

\bibitem{argus}
H.~Albrecht et. al.~(ARGUS Collaboration), Phys.\ Lett.\ {\bf B210}, 263
(1988).

\bibitem{cleo}
G.~Crawford et al.~(CLEO Collaboration), Phys.\ Rev.\ {\bf D45}, 752
(1992).

\bibitem{zoeller}
M.M.~Zoeller (CLEO Collaboration), Ph.D.~Thesis, submitted to the State
University of New York, Albany (1994). 

\bibitem{BHP}
T.E.~Browder, K.~Honscheid and S.~Playfer, Cornell Report No.~CLNS
93/1261, to appear in {\it $B$ Decays}, second edition, ed.~S.~Stone,
World Scientific.

\bibitem{simon}
A.~Simon, talk given at XXIX Rencontres de Moriond, "QCD and High Energy
Hadronic Interactions," March 1994, for the CERN WA89 hyperon beam  
experiment.

\bibitem{models}
I.I.~Bigi, Phys.\ Lett.\ {\bf 106B}, 510 (1981); V.L.~Chernyak and
I.R.~Zhitnitsky, Nucl.\ Phys.\ {\bf B345}, 137 (1990); P.~Ball and
H.G.~Dosch, Z.\ Phys.\ {\bf C51}, 445 (1991); M.~Jarfi et al., Phys.\
Rev.\ {\bf D43}, 1599 (1991).

\bibitem{ross}
W.R.~Ross, private communication for the CLEO Collaboration.

\bibitem{CD}
I.~Dunietz and P.S.~Cooper, Fermilab Report
No.~FERMILAB--PUB--94--107--T, in preparation.

\bibitem{browder}
T.E.~Browder, private communication for the CLEO Collaboration.

\bibitem{SVbaff}
I.I.~Bigi, B.~Blok, M.A.~Shifman and A.I.~Vainshtein, Phys.\ Lett.\ {\bf
B323}, 408 (1994).

\bibitem{DFW}
A.F.~Falk, M.B.~Wise and I.~Dunietz, Fermilab Report
No.~FERMILAB--PUB--94--106--T (1994).

\bibitem{argus2}
H.~Albrecht et al.~(ARGUS Collaboration), Z.\ Phys.\ {\bf C42}, 519
(1989).

\end{references}
\end{document}